\documentclass[]{spie}  

\usepackage{amsmath,amsfonts,amssymb}
\usepackage{graphicx}
\usepackage{booktabs}
\usepackage{siunitx}
\usepackage{subcaption}
\usepackage{url}
\usepackage[colorlinks=true, allcolors=blue]{hyperref}
\usepackage{authblk}
\usepackage[dvipsnames]{xcolor}
\usepackage[normalem]{ulem}

\title{Data-driven acquisition offset correction for telescope networks using machine learning and sensitivity analysis}

\authorinfo{Further author information: Send correspondence to J.Z.}

\begin{document}

\author[a,b]{Jackson Zariski}
\author[b]{Kaitlin M. Kratter}
\author[c]{Mark Bowman}
\author[c,d]{Nandini Dharwadkar}

\affil[a]{Department of Applied Mathematics, University of Arizona, Tucson, AZ, USA}
\affil[b]{Steward Observatory, University of Arizona, Tucson, AZ, USA}
\affil[c]{Las Cumbres Observatory, Santa Barbara, CA, USA}
\affil[d]{University of California, Santa Barbara, Santa Barbara, CA, USA}
\maketitle

\begin{abstract}
Reliable target acquisition is essential for maximizing observing efficiency, particularly for robotic telescope networks that operate across many sites, instruments, and environmental conditions. Although these facilities generate large volumes of operational and observational data, acquisition procedures often continue to rely on static pointing models and manual adjustments by observatory staff. These approaches can be effective, but they may not fully capture changing behavior associated with telescope configuration, observing conditions, instrument state, or other time-dependent effects. We present a data-driven machine-learning pipeline for automatically predicting and correcting telescope acquisition offsets, demonstrated using archival operational logs from the Las Cumbres Observatory network. Our approach applies both traditional machine-learning methods and deep-learning regression models to learn relationships between measured acquisition offsets and a range of observing, telescope, and instrumentation conditions. By using historical performance data, the pipeline is designed to complement existing pointing models and reduce the need for repeated manual correction. We also present an open-source, Dockerized software service that makes the trained models accessible through a browser-based graphical interface. The service is intended to support straightforward, real-time use by observatory staff while remaining portable across computing environments. Together, the prediction pipeline and deployment framework provide a practical path toward more adaptive, automated, and data-informed telescope acquisition.
\end{abstract}


\section{INTRODUCTION}

Modern astronomical facilities depend on large-aperture telescopes coupled to increasingly sensitive imaging and spectroscopic instruments. These systems are often required to acquire targets within tight tolerances, place them accurately on science detectors or entrance apertures, and maintain stable tracking over long integrations as the apparent sky motion is driven by Earth’s rotation. In this context, pointing refers to the accuracy with which the telescope slews to a commanded celestial position, while guiding refers to the closed-loop corrections used to hold the target at a desired detector location during an exposure. Achieving arcsecond-level acquisition accuracy is challenging because the delivered pointing depends not only on the nominal coordinate transformation from sky coordinates to mount coordinates, but also on mechanical flexure, encoder and alignment errors, atmospheric refraction, instrument configuration, thermal effects, and time-dependent behavior of the telescope structure. These effects become especially important for meter-class and larger facilities, where the moving mass of the telescope, mount, and attached instrumentation can be substantial, and where small angular errors correspond to meaningful offsets at the focal plane \cite{problempaper,astr-book}.

These sources of pointing error are well known to astronomers, and a variety of techniques have been developed to correct them. In many cases, however, existing pointing solutions rely on procedures that are computationally expensive, operationally inefficient, or difficult to update. Corrections for telescope motion may require manual intervention or extensive calculations, while mechanical effects are often represented using large polynomial models that estimate an offset to be applied to the commanded horizon coordinates. The most widely used proprietary software for this purpose is TPoint \cite{tpoint}. Constructing and maintaining these models can also require dedicated pointing runs, which consume observing time and may need to be repeated as telescope or instrument behavior changes.

A data-driven approach offers several practical advantages. Machine-learning models can be trained directly on acquisition measurements collected during normal observatory operations, reducing the need for specialized calibration campaigns and allowing the models to capture the conditions under which the telescope is actually used. Training and model updates can be performed automatically during daytime or other non observing periods, while newly collected data can be incorporated daily to account for changes in telescope alignment, instrumentation, or environmental conditions. This provides a path toward pointing corrections that improve over time without requiring repeated manual refitting. Motivated by these advantages, we train machine-learning models to recover and potentially improve upon existing telescope acquisition corrections.

Recent work has also explored machine-learning-based correction of telescope pointing errors. For example, Ref.~\cite{otherpaper} used historical South Pole Telescope observations and XGBoost models to predict azimuth and elevation control corrections from current weather conditions, and reported improved on-sky pointing performance after deployment. This provides an important proof of concept for data-driven pointing correction in telescope operations, while our work differs in telescope context, input feature set, prediction target, and software distribution.

 We include these tools in a GUI designed to be generalizable to a variety of telescope systems. We also provide the pointing pipeline as a Dockerized API, both with manual options and with a \texttt{flask} service. 
\section{DATA AND ACQUISITION PROBLEM}

The Las Cumbres Observatory (LCOGT) is a globally distributed, fully robotic telescope network that operates telescopes across multiple observatory sites as a single instrument, primarily for time-domain astronomy \cite{Brownlco}. For our purposes, we consider data from LCOGT's South African node at Sutherland, which hosts multiple LCOGT telescopes. Unlike long time-series observations of a small number of targets, the LCOGT sample considered here consists of many individual robotic acquisitions drawn from network operations and pointing-calibration sequences.

For each observation we are given the observation date, the local sidereal time, the mount-commanded right ascension and declination, and the astrometrically measured sky coordinates from the image world coordinate system (WCS), expressed in a common frame. We refer to the demanded telescope coordinates as TPT coordinates. The target in this problem is the acquisition offset between the demanded TPT coordinate and the WCS coordinate returned from the astrometric solve.

A typical data log from the observatory contains the following information:
\begin{itemize}
    \item the time and date of the observation, broken into year, month, day, hour, minute, and second;
    \item TPT right ascension and TPT declination;
    \item WCS right ascension and WCS declination.
\end{itemize}
The LCOGT system also has enough latency between acquisitions that error in the previous acquisition can be calculated and passed through as a feature. In training we initialize previous acquisition error as $(0,0)$, and the previous-error features at label $i$ are the target offset values from index $i-1$.

\section{SYNTHETIC LCOGT EXPERIMENTS}

When approaching a supervised-learning problem like acquisition offset correction, choosing an appropriate model architecture is crucial. While large, densely connected neural networks can capture detailed nonlinearities if tuned properly, certain studies have also shown that XGBoost performs well on physics-based problems with limited training data. With the ultimate goal of this work being the creation of a general framework for acquisition software, it is helpful to have concrete metrics to indicate that one architecture is better suited to the physical situation over another. In addition, while it is always possible to include many features in a model and hope training will tune irrelevant weights to zero, in larger dimensions this can be costly in both time and memory complexity and is not always guaranteed to succeed. Feature engineering through informed variable choice prior to training is preferred.

These informed choices on architectures and features rely on experimental statistical techniques, using what we call synthetic data: data that we design to closely model what we expect the real system to look like, based on known physical properties of the system, measurement noise, or empirical distributions. To frame the LCOGT investigation, we begin with synthetic test cases modeled on real data taken from the aforementioned LCOGT telescope in South Africa.

A natural question is whether acquisition architectures, either deep or tree-based, can capture the raw transformations from celestial to horizon coordinates that we normally delegate to packages such as \texttt{AstroPy} or \texttt{PyEphem} \cite{astropy:2013,pyephem}. Using a set of 14,589 observations and a 70/10/20 split for training, validation, and testing with an XGBoost model, a simple experiment to replicate the function
\begin{equation}
    f:[\mathrm{LST},\mathrm{RA},\mathrm{Dec}]\to[\mathrm{Az},\mathrm{El}]
\end{equation}
using the WCS segment of data results in Table~\ref{tab:xgbtrans}.

\begin{table}[ht]
\caption{Astropy transformation XGBoost results.}
\label{tab:xgbtrans}
\begin{center}
\begin{tabular}{lcc}
\toprule
 & Azimuth & Elevation \\
\midrule
Mean absolute error (degrees) & 0.372 & 0.138 \\
Median absolute error (degrees) & 0.152 & 0.079 \\
\bottomrule
\end{tabular}
\end{center}
\end{table}

\begin{table}[ht]
\caption{Astropy transformation neural-network results.}
\label{tab:nntrans}
\begin{center}
\begin{tabular}{lcc}
\toprule
 & Azimuth & Elevation \\
\midrule
Mean absolute error (degrees) & 0.789 & 0.138 \\
Median absolute error (degrees) & 0.601 & 0.109 \\
\bottomrule
\end{tabular}
\end{center}
\end{table}

\begin{figure}[htbp]
  \centering
  \begin{subfigure}{0.48\linewidth}
    \centering
    \includegraphics[width=\linewidth]{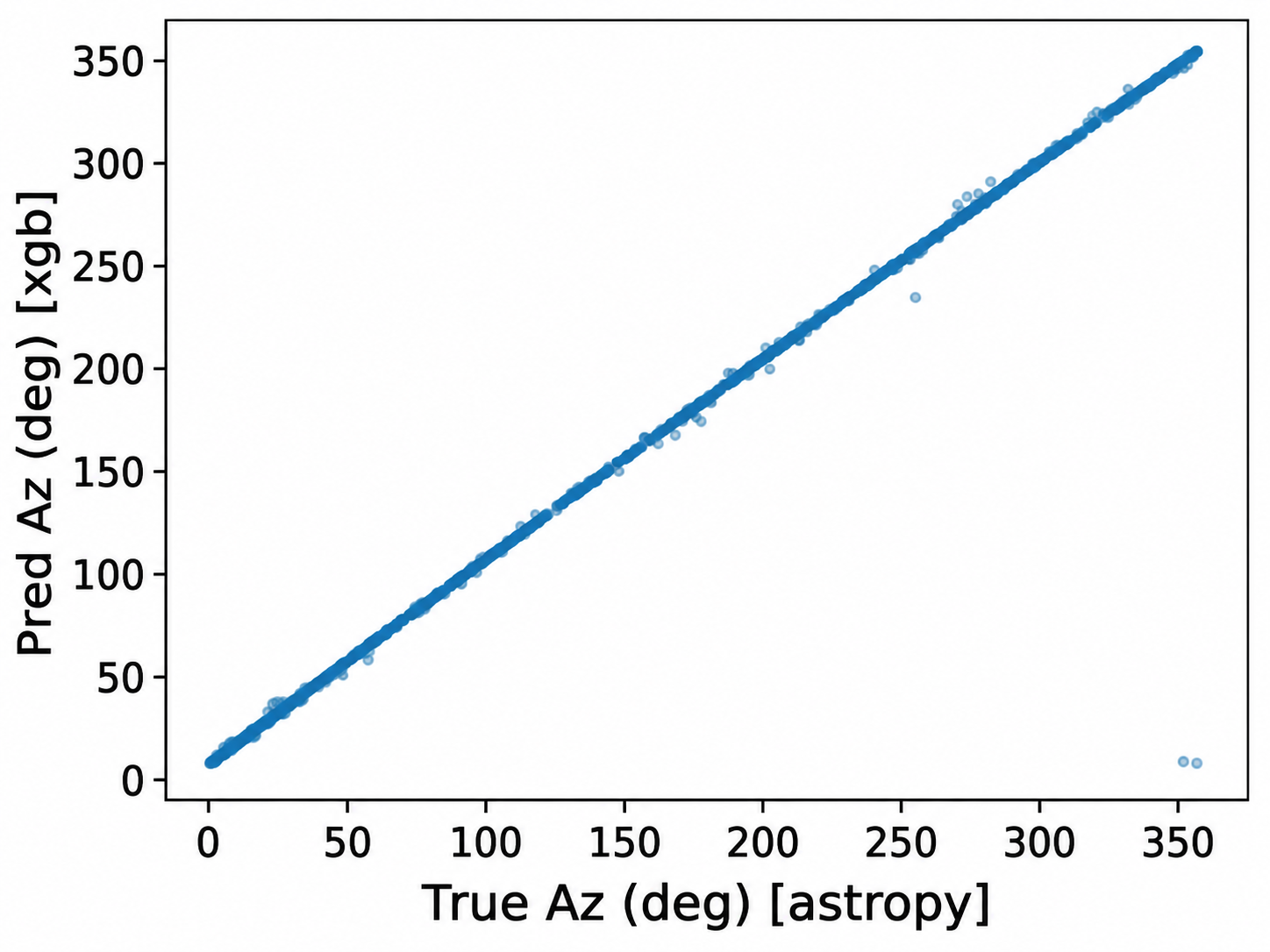}
  \end{subfigure}\hfill
  \begin{subfigure}{0.48\linewidth}
    \centering
    \includegraphics[width=\linewidth]{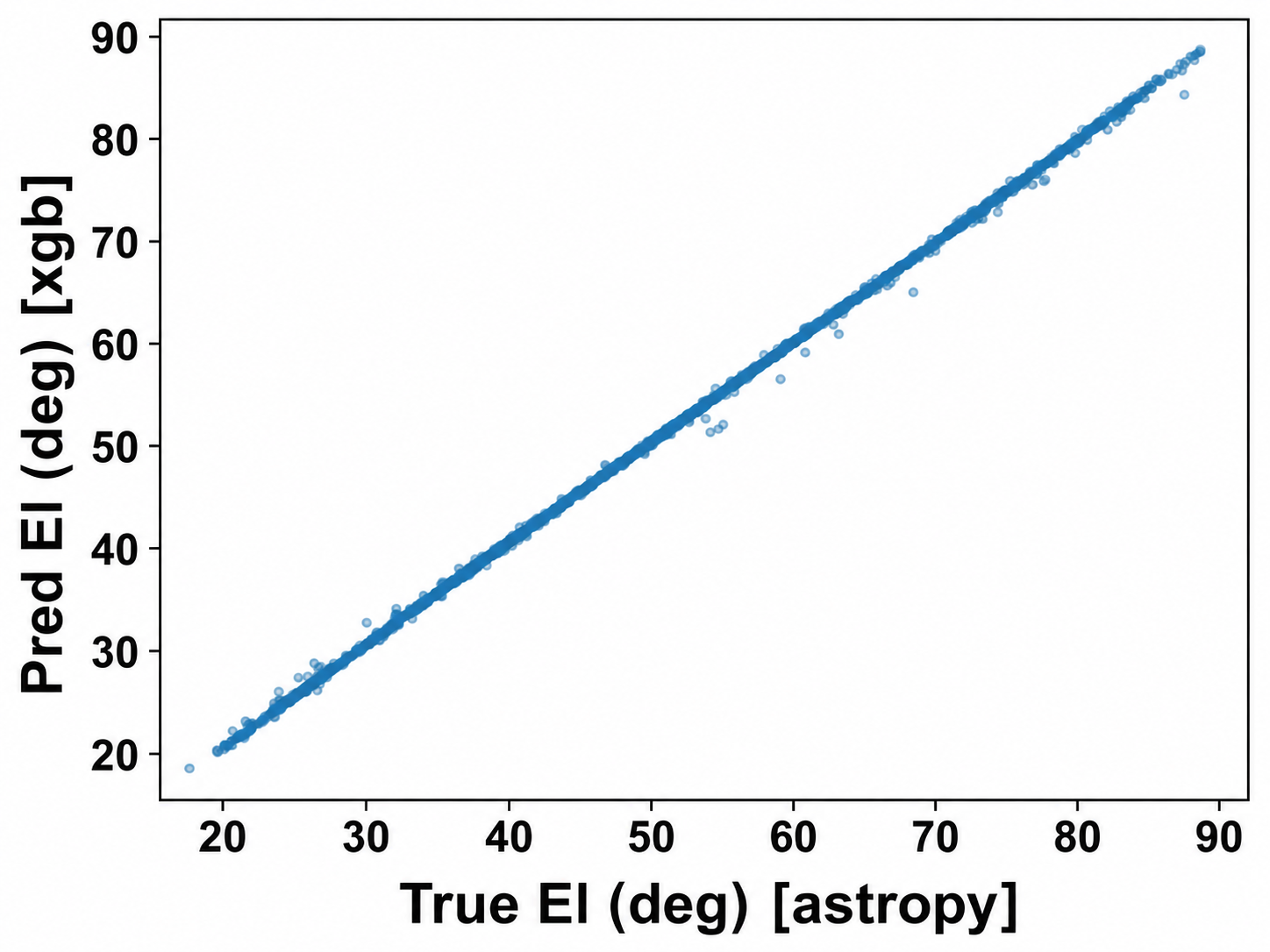}
  \end{subfigure}
  \caption{Parity results for the coordinate-conversion replication test.}
  \label{fig:paritygraph}
\end{figure}

Figure~\ref{fig:paritygraph} demonstrates that the basic premise of the model is sound. The network can recover the main astronomical transformation itself, i.e. the nonlinear spherical-trigonometric relation between sky coordinates, time, and local telescope coordinates. This is important, since any successful acquisition model that lacks these initial features must first reproduce this dominant geometric structure before it can hope to learn the smaller residual effects present in real data.

The next question is whether the model can also capture typical perturbations and systematic offsets superimposed on the baseline coordinate relation. One example is a zero-point offset, which appears as an approximately constant positional bias--the mount encoder may report positions $(A_i,h_i)$ in degrees, while the coordinate solution from the image returns a consistently displaced location
\begin{equation}
    (A_i-Z_A,h_i-Z_h),\qquad Z_A,Z_h\in\mathbb{R}.
\end{equation}
In our simplified test case we represent additional systematic perturbations as sinusoids,
\begin{equation}
A^i_{\mathrm{sys}} = C_1\sin(A^i) + C_2\cos(h^i),
\end{equation}
\begin{equation}
h^i_{\mathrm{sys}} = C_3\cos(A^i) + C_4\sin(h^i),
\end{equation}
where all $C_i$ are constants tuned to the order of arcminutes. To mimic real observing conditions, we add Gaussian distributed noise with $\mu=0$ on the order of a few arcseconds, denoted $(\epsilon_{A^i},\epsilon_{h^i})$. Thus the synthetic target is
\begin{equation}
 f(A^i,h^i)=
 \begin{bmatrix}
    A^i_{\mathrm{target}}\\
    h^i_{\mathrm{target}}
\end{bmatrix}
=
\begin{bmatrix}
    A^i-Z_A+A^i_{\mathrm{sys}}+\epsilon_{A^i} \\
    h^i-Z_h+h^i_{\mathrm{sys}}+\epsilon_{h^i}
\end{bmatrix}.
\end{equation}

\begin{figure}[htbp]
    \centering
    \includegraphics[width=0.72\textwidth]{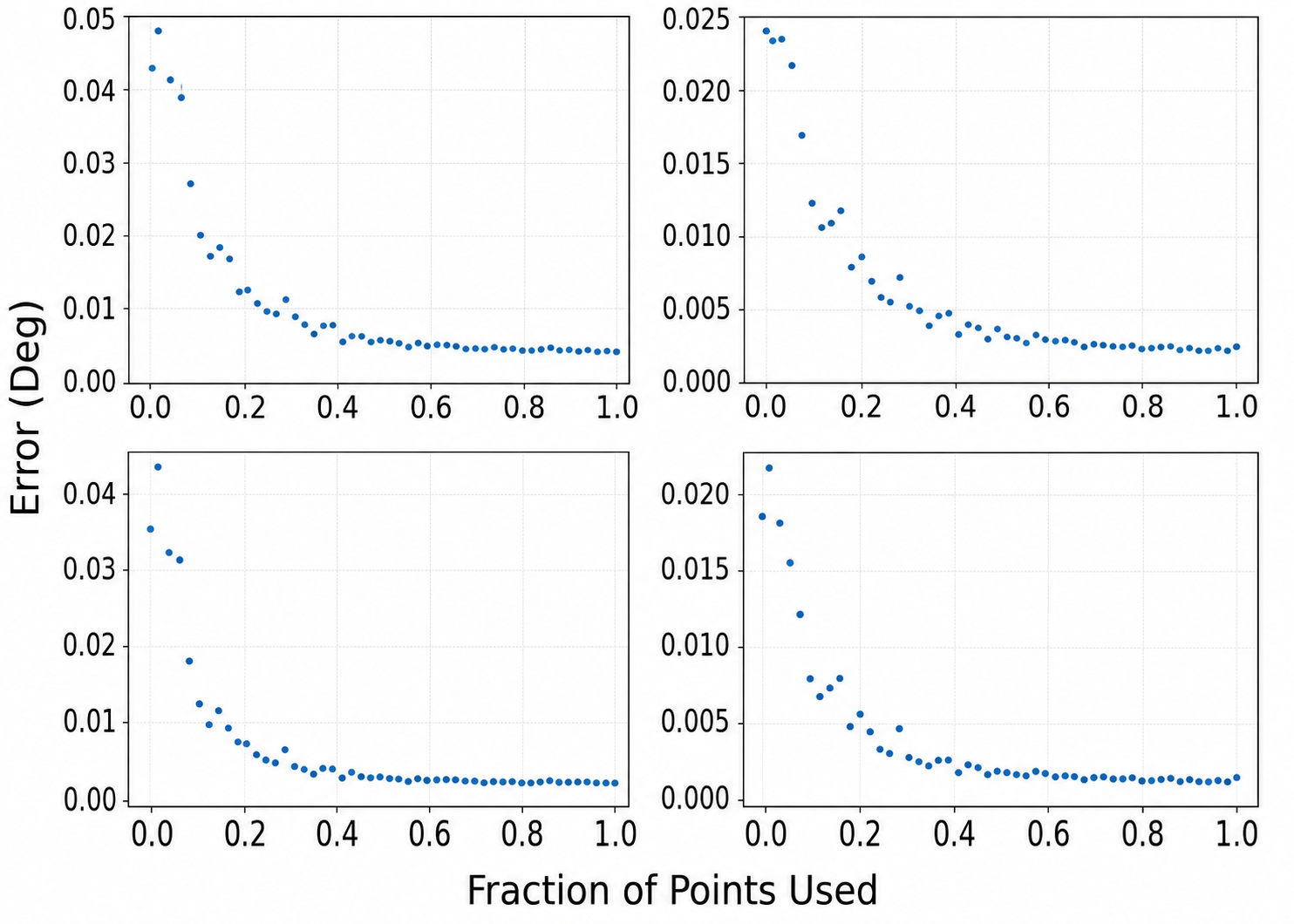}
    \caption{Accuracy convergence in the synthetic-data test. XGBoost rapidly coalesces to a steady accuracy score as we approach using the roughly 10,000 points available in this data log.}
    \label{fig:covplot1}
\end{figure}

Using the same features described previously, we train a model to predict $(A^i_{\mathrm{target}},h^i_{\mathrm{target}})$. Figure~\ref{fig:covplot1} displays coalescence of model error to the order of induced noise as we increase the fraction of training points used. Most of the improvement occurs within the first 30–40 percent of the available data, after which the errors decrease more gradually and approach the scale set by the injected Gaussian noise. For each train-test run, we hold the zero-point offsets, sinusoidal perturbations, noise realization, model architecture, and test set fixed, so that the amount of training data is the only quantity varied. Figure~\ref{fig:covplot1} therefore demonstrates both the model’s ability to recover the imposed systematic structure and the diminishing improvement obtained from adding further training examples.

\begin{figure}[htbp]
    \centering
    \includegraphics[width=0.72\textwidth]{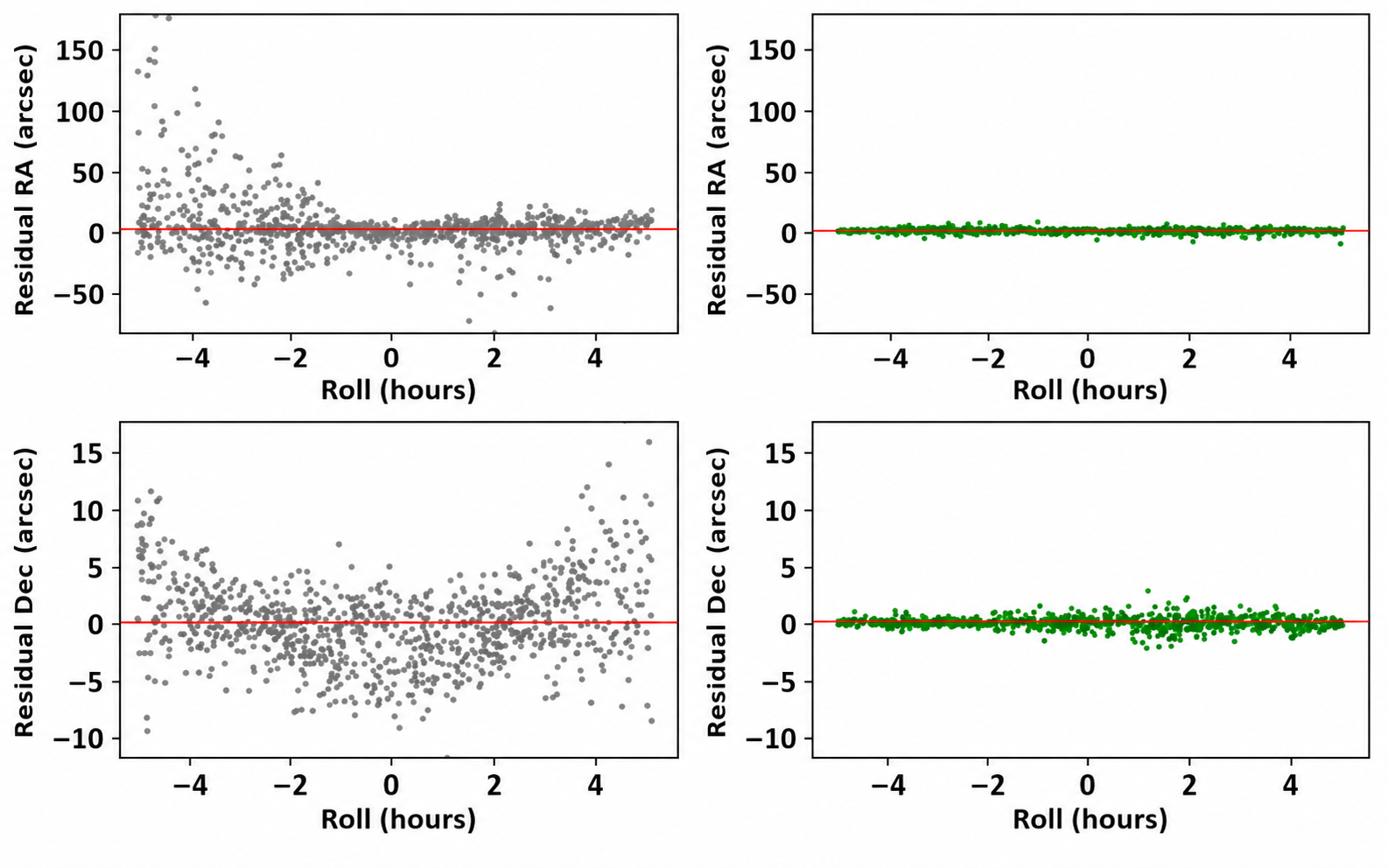}
\caption{Residuals from XGBoost models trained on synthetic acquisition offsets based on the roll- and pitch-dependent structure of an existing LCOGT TPoint model. The upper panels show right-ascension residuals as a function of roll, while the lower panels show declination residuals. The left panels use the raw angular coordinates as features, whereas the right panels additionally include their sine and cosine transformations. The trigonometric encoding accounts for the periodicity of the mount coordinates and produces residuals that are more tightly concentrated around zero.}
    \label{fig:res1}
\end{figure}

Finally, to construct a more physically motivated synthetic test, we base the systematic offset terms on the roll- and pitch-dependent behavior of an existing LCOGT TPoint solution. Because roll and pitch are angular variables, models trained only on their raw numerical values do not naturally represent the equivalence of values near the coordinate-wrapping boundary. We therefore augment each angular feature with its sine and cosine components and compare the resulting residuals with those obtained using the raw coordinates alone. The importance of this cyclical encoding is displayed in Figure \ref{fig:res1}, where we see that the inclusion of said feature engineering reduces model dependency on artificial coordinate boundaries and de-trends the systems to more uniform accuracy.

\section{LCOGT RESULTS ON HISTORICAL TELEMETRY}

With the synthetic experimentation and variance analysis studies for architecture choice and feature engineering completed, we proceed with the true LCOGT problem. Here, we seek to use the provided TPT celestial coordinates along with other features to predict the offsets of the WCS solution. TPT represents the coordinate demand sent to the telescope system. With previous acquisition error included as an additional feature, we proceed with a traditional extraction and transformation data pipeline, separating training and testing sets temporally, and evaluating the resulting model.

\begin{figure}[htbp]
    \centering
    \includegraphics[width=0.72\textwidth]{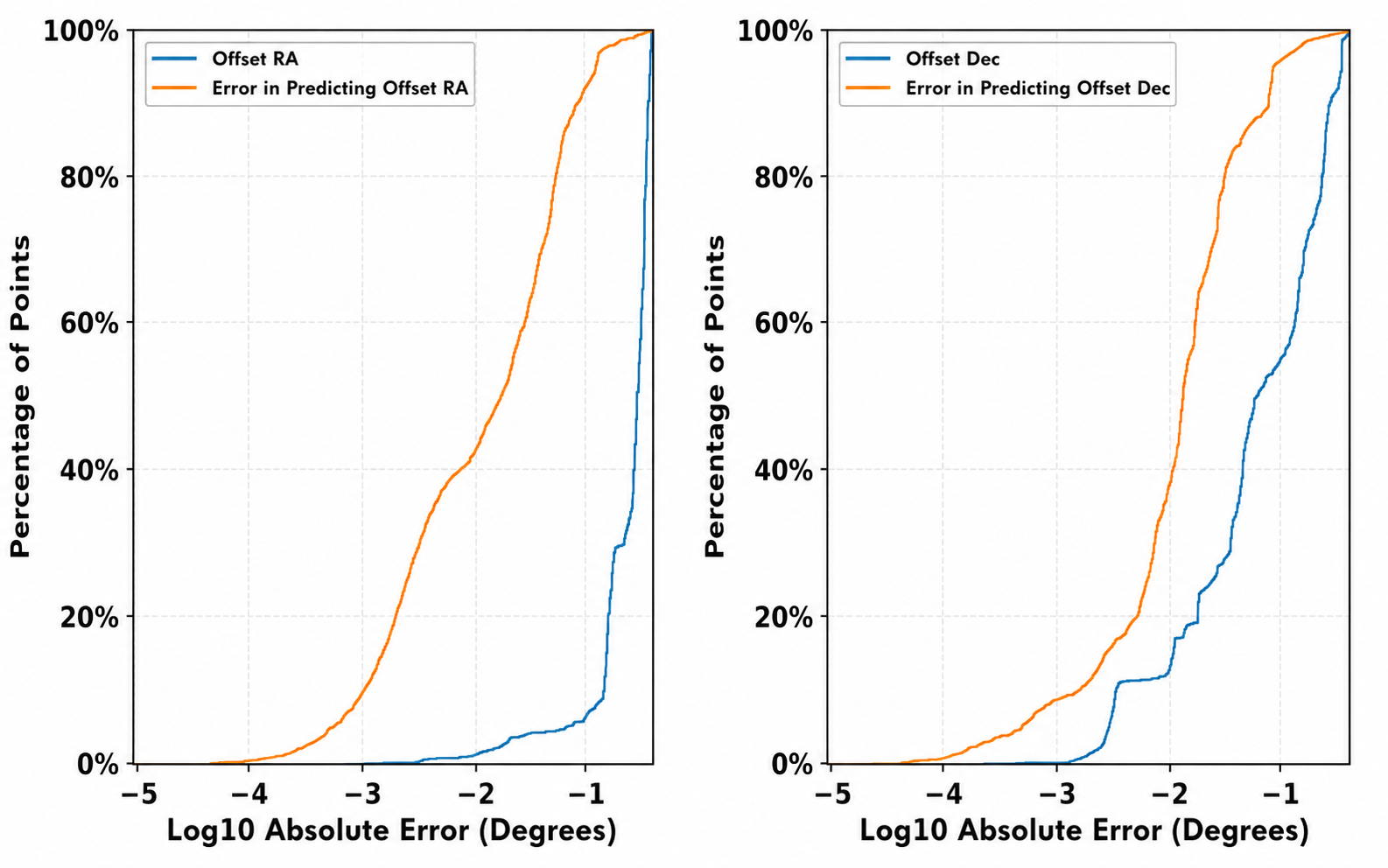}
    \caption{Empirical cumulative distributions of the log-transformed absolute acquisition-offset errors on the temporally held-out LCOGT test set in right ascension (left) and declination (right). Each panel compares the original offsets with the residual errors after applying the XGBoost-predicted corrections. The corrected-error curves are shifted toward smaller values, indicating that a larger fraction of test observations falls below any given error threshold.}
    \label{fig:truecdfs}
\end{figure}

Figure~\ref{fig:truecdfs} shows the results of true data experiments. In Tables~\ref{tab:lco_real_ra} and \ref{tab:lco_real_dec}, the gradient-boosted tree model reduces mean and median offset error by roughly an order of magnitude in both coordinate directions. Hence, we conclude that the system is a viable alternative to traditional acquisition correction techniques in the LCOGT case, with next steps being a true test on-sky.

\begin{table}[htbp]
\centering
\caption{Results in RA space with real LCOGT acquisition data logs.}
\label{tab:lco_real_ra}
\begin{tabular}{lccc}
\toprule
Degrees & Mean & Median & 90th percentile \\
\midrule
Original offset RA error & 0.354 & 0.372 & 0.516 \\
Corrected model RA error & 0.0323 & 0.0164 & 0.0836 \\
\bottomrule
\end{tabular}
\end{table}

\begin{table}[htbp]
\centering
\caption{Results in declination space with real LCOGT acquisition data logs.}
\label{tab:lco_real_dec}
\begin{tabular}{lccc}
\toprule
Degrees & Mean & Median & 90th percentile \\
\midrule
Original offset Dec error & 0.059 & 0.0447 & 0.161 \\
Corrected model Dec error & 0.012 & 0.00698 & 0.0333 \\
\bottomrule
\end{tabular}
\end{table}

For programming the neural networks, we used \texttt{TensorFlow}, which allows for the construction of networks with back-propagation done internally on the back end \cite{tensorflow2015-whitepaper}. For the gradient-boosted tree model, we used the \texttt{XGBoost} package in Python \cite{xgboost}. We have validated that the model functions in reverse, by using celestial coordinates to predict mount-encoder horizon coordinates, but cannot yet generate a robust comparison with the existing pointing model in this direction.

\section{DOCKERIZED API AND FLI-POINT GUI}

One of the key goals for this project was the development of shareable software that could accomplish two objectives. First, the software needs to be more accurate than what is currently available; an inferior open-source product is not useful for an observatory to integrate into its framework. Second, the software must generalize to a variety of systems and operators. Some operators may want a function they can inject directly into their system; others, such as LCOGT, prefer a Dockerized application programming interface (API). The API is an effective means of distribution because it can be run out of \texttt{bash} without needing prior machine-learning knowledge. In addition to these software packages, another way to ensure ease of distribution is to package the system in an easy-to-use graphical user interface. Written with \texttt{streamlit}, we name this GUI the FLI-Point, or Fast-Learning Informed Pointing, system \cite{streamlit}.

For ease of use, we provide two different API systems. The first API is a direct copy of the training pipeline. After providing the pre-processed data file, the user runs each script in order, with Sobol feature analysis optional. The second version of this API wraps the same system with a \texttt{flask} container, providing methods for interacting with other machines and an interface for injection into existing telescope pointing systems. The \texttt{flask} version provides \texttt{train} and \texttt{predict} methods that can interact with JSON files and remote apps from telescope systems. Both APIs are designed to aid the acquisition problem, though work on tracking software is underway.

All data are expected to be pre-processed as a whitespace-delimited text file. The first six columns are expected to follow the structure in Table~\ref{tab:api_requirements}. Additionally, longitude, latitude, and elevation values are needed for conversion from celestial offsets to the horizon system.

\begin{table}[htbp]
\centering
\caption{Processing requirements for the pointing API.}
\label{tab:api_requirements}
\begin{tabular}{ll}
\toprule
Column & Meaning \\
\midrule
DATE-OBS & ISO date/time, e.g., 2025-05-09T04:26:12.247 \\
LST & Sidereal time, HH:MM:SS[.sss] \\
OBS-RA & Pointing RA in hours, HH:MM:SS[.sss] \\
OBS-DEC & Pointing Dec in degrees, $\pm$DD:MM:SS[.ss] \\
CRVAL1 & Solved RA in degrees \\
CRVAL2 & Solved Dec in degrees \\
\bottomrule
\end{tabular}
\end{table}

In addition to the pair of APIs, FLI-Point provides a graphical user interface to simplify the training and prediction process for acquisition. A data file must be formulated in the same fashion as the Docker service, but the user is allowed some freedom in training choice. These choices include the type of model, the learning rate, the number of estimators or hidden-layer sizes, the maximum tree depth or number of hidden layers, early stopping rounds, optional feature engineering, temporal splitting, train/evaluation/test percentages, and whether to generate additional statistics and plots on the testing set.

\begin{figure}[htbp]
  \centering
  \begin{subfigure}{0.82\linewidth}
    \centering
    \includegraphics[width=\linewidth]{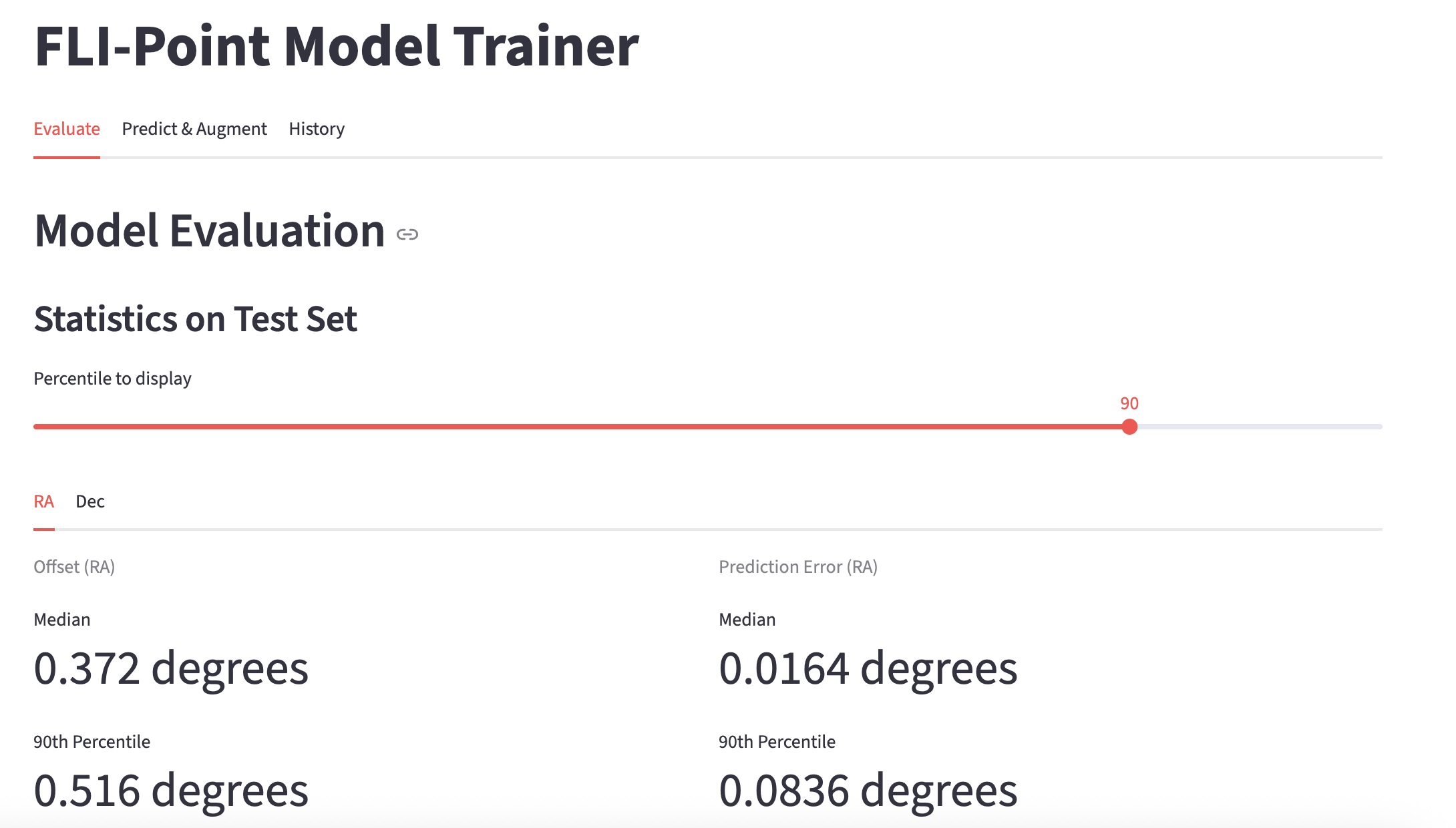}
  \end{subfigure}

  \vspace{0.5em}

  \begin{subfigure}{0.82\linewidth}
    \centering
    \includegraphics[width=\linewidth]{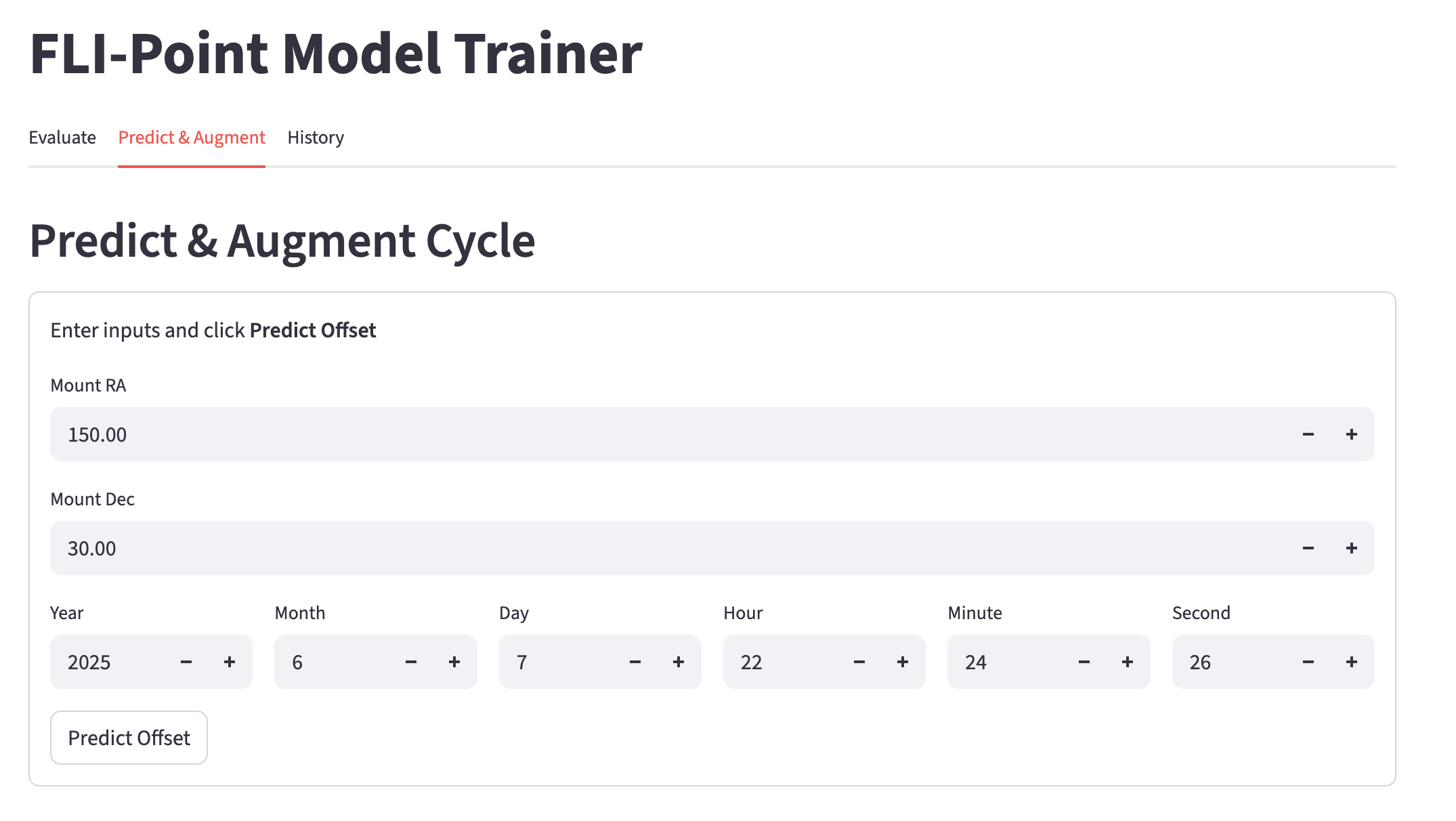}
  \end{subfigure}

  \caption{Example pages of the FLI-Point interface, showing a testing-results page and an entry point for predictions. The user is also able to adjust which percentile of statistics to view.}
  \label{fig:flipoint}
\end{figure}

Beyond allowing users to customize training hyperparameters, FLI-Point also supports the post-training prediction workflow. Users can enter desired celestial pointing coordinates, the expected acquisition time, and any additional required features to obtain a predicted offset correction. The interface can also be used for shadow-mode testing, in which the machine-learning model generates and records predicted corrections during an observing run while the telescope continues to operate using its native pointing software. These predictions can then be compared with the measured acquisition offsets without affecting telescope control, providing a low-risk way to evaluate model performance under real observing conditions before integrating it into the operational correction loop. Figure~\ref{fig:flipoint} shows the model-evaluation and prediction pages used for these tasks.

\section{CONCLUSION}

Modern observatories generate an overwhelming amount of telemetry, yet acquisition and pointing workflows still often depend on static pointing models and hand-tuned heuristics that are updated infrequently. We have shown that we can do better by treating acquisition logs as not only a byproduct of scientific observations, but as something we can mine, model, interpret, and deploy back into regular pointing operations. The main result is a unified framework for predicting and correcting on-sky acquisition offsets using supervised machine learning. We paired this with interpretation tools that make the models usable by a general operator, and wrapped them in software that is practical for observatory environments.

On the modeling side, we showed that archival acquisition records often, under the correct conditions, contain enough information to learn a meaningful mapping from sky coordinates and other features to expected acquisition offsets. By training regression models on historical logs, we demonstrated that data-driven predictors can substantially reduce theoretical offset errors compared to an uncorrected baseline without demanding additional observing time. While no single model generalizes perfectly across all telescopes, the consistent pattern is that the general system contains repeatable, learnable structure in its errors, and that this structure can be exploited by both traditional and deep supervised learning methods.

The immediate next step for the LCOGT work is a true on-sky test, first in shadow mode and then, if successful, as part of an acquisition correction loop. The broader goal is a service that can be retrained regularly on normal operational telemetry and used by observatory staff without requiring machine-learning expertise.

\bibliographystyle{spiebib}
\bibliography{spie_2026_references}

\end{document}